\documentclass[pra,showpacs,graphicx,multicols,amsmath,amssymb,floatfix]{revtex4}

\usepackage{amssymb}
\usepackage[dvips]{color}
\usepackage{newlfont}
\usepackage{amssymb}
\usepackage{amsfonts}
\usepackage{amsmath}
\usepackage{graphicx}
\usepackage{psfrag}
\usepackage{bm}

\newcommand{\matwo}[1]{\left(\begin{array}{cc} #1 \end{array}\right)}

\newcommand{\ket}[1]{|#1\rangle}

\newcommand{\ha}{\hat a}
\newcommand{\hb}{\hat b}
\newcommand{\hn}{\hat n}
\newcommand{\hN}{\hat N}
\newcommand{\hL}{\hat L}
\newcommand{\hO}{\hat O}
\newcommand{\hH}{\hat H}
\newcommand{\hrho}{\hat\rho}
\newcommand{\had}{\ensuremath{\ha^\dagger}}
\newcommand{\hbd}{\ensuremath{\hb^\dagger}}
\newcommand{\gnl}{{\cal G}}
\newcommand{\gl}{{\cal U}}

\begin{document}

\title{Collisional shifts in optical-lattice atom clocks}

\author{Y. B. Band$^{1,2}$ and A. Vardi$^1$}
\affiliation{$^1$ Department of Chemistry, Ben-Gurion University of the Negev,
Beer-Sheva 84105, Israel\\
$^2$ Atomic Physics Division, A267 Physics, National Institute of
Standards and Technology, Gaithersburg, MD 20899 and The Ilse Katz
Center for Nano-Science, Ben-Gurion University of the Negev,
Beer-Sheva 84105, Israel}

\begin{abstract}
We theoretically study the effects of elastic collisions on the
determination of frequency standards via Ramsey fringe spectroscopy in
optical-lattice atom clocks.  Interparticle interactions of bosonic
atoms in multiply-occupied lattice sites can cause a linear frequency
shift, as well as generate asymmetric Ramsey fringe patterns and
reduce fringe visibility due to interparticle entanglement.  We
propose a method of reducing these collisional effects in an optical
lattice by introducing a phase difference of $\pi$ between the Ramsey
driving fields in adjacent sites.  This configuration suppresses site
to site hopping due to interference of two tunneling pathways, without
degrading fringe visibility.  Consequently, the probability of double
occupancy is reduced, leading to cancellation of collisional shifts.
\end{abstract}


\maketitle

\section{Introduction}

Current state-of-the-art atomic clock technology is based mainly on
trapped single ions or on clouds of free falling cold atoms
\cite{clock_reviews}.  Recently, a new type of atomic clock based on
neutral atoms trapped in a deep ``magic-wavelength'' optical lattice,
wherein the ground and excited clock states have equal light shifts,
has been suggested \cite{Takamoto_03, Katori_03}.  For interfering,
far-detuned light fields, giving rise to a ``washboard'' intensity
pattern, the optical potential experienced by an atom in a given
internal state is $V_0({\mathbf r}) = \hbar |\Omega_0({\mathbf
r})|^{2}/(4\Delta_0)$ where $\Omega_0({\mathbf r}) = 2\mu E_0({\mathbf
r})/\hbar$ is the dipole coupling frequency, $E_0({\mathbf r})$ is the
radiation electric field strength at position ${\mathbf r}$, and
$\Delta_0$ is the detuning of the oscillating optical field from
resonance.  In a magic-wavelength optical lattice, the same exact
potential is experienced by the ground and the excited state of the
clock transition.

Optical-lattice atomic clocks are advantageous due to the suppression
of Doppler shifts by freezing of translational motion (the clock
operates in the Lamb-Dicke regime, where atoms are restricted to a
length-scale smaller than the transition wavelength \cite{Dicke_53}),
the narrow linewidth due to the long lifetime of the states involved
in the clock transition, and the large number $M$ of occupied sites,
minimizing the Allan standard deviation \cite{clock_reviews}.  As
already mentioned, the optical-lattice potential light shift is
overcome by using trapping lasers at the magic wavelength
\cite{Katori_03, Takamoto_03}.  Operating optical-lattice clocks in
the optical frequency regime, rather than in the microwave, has the
added benefit of increasing the clock frequency $\nu$ and thereby
reducing $\delta \nu/\nu$ by five orders of magnitude.  A coherent
link between optical frequencies and the microwave cesium standard is
provided by the frequency comb method \cite{comb}.  Thanks to these
characteristics, atomic optical-lattice clocks have promise of being
the frequency standards of the future.  Attempts to further improve
the accuracy of such clocks using electromagnetically induced
transparency to obtain accuracies on the order of $10^{17}$ or better
have also been suggested \cite{Santra_05}.

A 3-dimensional (3D) optical lattice configuration would allow the
largest number of atoms that can participate in an optical-lattice
atom clock.  However, when sites begin to be multiply-occupied,
atom-atom interactions can shift the clock transition frequency
\cite{inter_shift}.  This is particularly problematic in a very deep
optical lattice since the effective density in sites with more than
one atom will then be very high due to the highly restricted volume.
Hence, the collisional shift, proportional to the particle density,
can be very large.  It is therefore important to ensure that atoms
individually occupy lattice sites (preferably in the ground motional
state of the optical lattice).  One way of achieving this goal is by
low filling of the optical-lattice, so that the probability of having
more than one atom per site is small.  If hopping of an atom into a
filled lattice site during the operation of the clock is small,
collisional effects will be minimal.

One kind of optical lattice clock configuration that can have minimum
collisional effects uses ultracold spin-polarized fermions in a deep
optical lattice.  For example, a system of ultracold optically pumped
$^{87}$Sr atoms in the 5s$^2$ $^1$S$_0$ $|F = 9/2,M_F= 9/2 \rangle$
internal state, filled to unit filling.  The nuclear spin of the atoms
is protected and one expects that the gas will remain spin polarized
for a very long time.  The probability of finding two fermionic atoms
in a deep lattice site is extremely low in such a system as long as
the gas is sufficiently cold, since higher occupied bands will not be
populated, and only one fermionic atom will be present in a given site
due to the Pauli exclusion principle.  Moreover, fermionic atoms
cannot interact via $s$-wave collisions and $p$-wave and higher
partial waves are frozen out at low collision energies.  Provided the
ground motional state of the spin polarized fermion system can be
attained by adiabatically turning-on the optical lattice
\cite{Julienne_05}, this system seems to be extremely well-suited for
making an accurate clock.

Another potentially interesting configuration involves ultracold
bosonic atoms such as $^{88}$Sr atoms in the 5s$^2$ $^1$S$_0$ $|F =
0,M_F= 0 \rangle$ internal state in an optical lattice with very low
filling.  Since the atoms are bosons, there is no
single-atom-occupancy constraint, allowing for collisional shifts.
These shifts can be minimized if the filling factor $p$ is small.
Moreover, multiple occupancy caused by tunneling between adjacent
populated sites is reduced if the optical lattice is deep and the
probability of hopping of an atom into an adjacent filled site,
$J_{\mathrm{hop}}$, is small.  Therefore, we expect that the
collisional shift should be very low.  Here, we investigate the Ramsey
fringe clock dynamics for such a system.  The transition between the
$^1$S$_0 \, |F = 0,M_F= 0 \rangle$ and the $^3$P$_0 \, |F = 0,M_F= 0
\rangle$ state of $^{88}$Sr can be a Raman transition
\cite{Santra_05}, but we can think of the transition between $^1$S$_0$
and $^3$P$_0$ as a Rabi transition as long as the detuning from the
intermediate state coupling these two levels is large.  Yet another
possibility is to use a weak static magnetic field to enable direct
optical excitation of forbidden electric-dipole transitions that are
otherwise prohibitively weak by mixing the $^3$P$_1$ with the
$^3$P$_0$ state \cite{Hollberg_06}.  In contrast to multiphoton
methods proposed for the even isotopes \cite{Santra_05}, this method
of direct excitation requires only a single probe laser and no
multiphoton frequency-mixing schemes, so it can be readily implemented
in existing lattice clock experiments \cite{Barber_06}.  I.e., this
method for using the ``metrologically superior'' even isotope can be
easily implemented.  However, one of the problems that can arise is
that more than one bosonic atom can fill a lattice site, and these
atoms can interact strongly.  We shall assume that sites are initially
populated with at most one atom per site, but that during the
operation of the Ramsey separated field cycle, i.e., the delay time
between the two $\pi/2$ Ramsey pulses, $T$, atoms from adjacent sites
can hop; if a site ends up with more than one atom during the cycle,
the clock frequency will be adversely affected by the collisional
shift.

It is easy to obtain an order of magnitude estimate for the
collisional shift $\delta \nu$ in this kind of clock.  The product of
the hopping rate $J_{\mathrm{hop}}$ and the Ramsey delay time $T$
gives the probability for hopping between sites.  As will be shown
below, the shift obtained is therefore given by $\delta \nu =
J_{\mathrm{hop}} T \gnl$, where $\gnl$ is the nonlinear interaction
strength parameter defined in Eq.~(\ref{GNL}).  If we consider only a
single site, then to attain an accuracy of one part in $10^{17}$, one
must have $J_{\mathrm{hop}} T \gnl \leq 10^{-17} \nu$, whereas for
$N_s$ occupied sites, $J_{\mathrm{hop}} T \gnl/\sqrt{N_s} \leq
10^{-17} \nu$.  The precise values of these parameters can vary
greatly for different experimental realizations.  In particular, the
hopping rate $J_{\mathrm{hop}}$ is exponentially dependent on the
optical lattice barrier height and on particle mass.

Three distinct collisional effects are found for a single,
multiply-populated lattice site.  These include a simple linear
frequency shift, a nonlinear shift resulting in asymmetric fringe
patterns, and an entanglement-induced reduction in fringe contrast.
In order to show how these effects can take place dynamically during
the application of the Ramsey pulse sequence, we consider a 1D optical
lattice and focus on two adjacent sites filled with one atom in each
site, and calculate the probability of double occupancy due to hopping
(tunneling) of an atom to its adjacent site and the resulting
collisional shifts.  We show that by varying the direction of the Rabi
drive laser with respect to the principal optical lattice axis, one
can induce a phase-difference between the Rabi drive fields in
adjacent lattice site.  We find that due to interference of tunneling
pathways, hopping is suppressed when this phase difference is set to
$\pi$, as compared to the case where all sites are driven in phase.
Consequently, detrimental collisional effects are diminished for
inverted-phase driving of adjacent sites.

\begin{figure}
\centering
\includegraphics[scale=0.6]{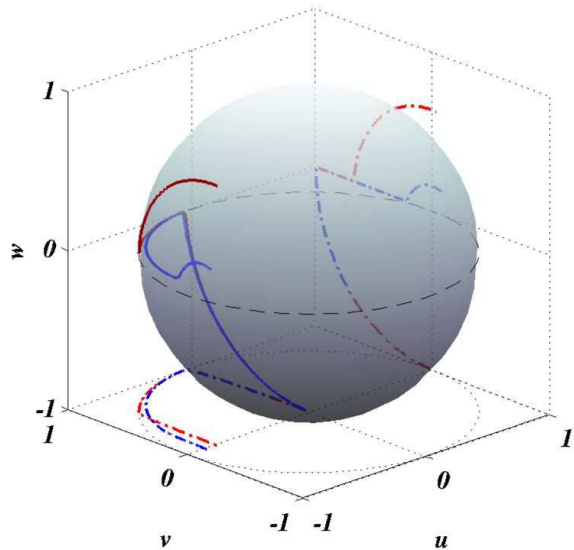}
\caption{(color online) Collisional shift and loss of single-particle
coherence in a Ramsey scheme.  The solid red line traces the
trajectory of the Bloch vector in the absence of interactions.  When
interactions are present (solid blue line), the frequency of phase
oscillations is modified if $G_{gg}\neq G_{ee}$ and the length of the
Bloch vector is not conserved when $G_{gg}+G_{ee}\neq 2G_{ge}$ (note
that the solid blue curve deviates from the dashed great circle during
the free induction decay, whereas the solid red curve does not).}
\label{fig1}
\end{figure}

The outline of the paper is a follows.  In Sec.~\ref{section2} we review
the Ramsey separated oscillatory fields method for noninteracting
particles and set up the notation we use in what follows.  
In Sec.~\ref{section3} we construct a second-quantized model,
describing the single-site Ramsey scheme for interacting bosons.
This model is numerically solved in Sec.~\ref{section4}, demonstrating
various collisional effects on Ramsey fringe patterns. Hopping between
sites is introduced in Sec.~\ref{section5} where we present two-site
results and propose an inverted Rabi phase scheme to cancel collisional
shifts. Conclusions are presented in Sec.~\ref{conclusion}.

\section{Ramsey separated oscillatory fields} \label{section2}

Norman Ramsey introduced the method of separated oscillatory fields in
1950 \cite{Ramsey_50}.  A long time-period between the application of two
nearly resonant coherent fields makes the Ramsey resonance very
narrow, and thus suitable for high-performance clocks and precision
measurements \cite{Vanier_89,Lemonde_01}.  The method has since become a
widely used technique for determining resonance frequencies to high
precision.  For example, in the Cs fountain clock experiments
summarized in Refs.~\cite{Lemonde_01,Bauch_03}, the observed linewidth
of the Ramsey resonance was 1 Hz, two orders of magnitude below that
of thermal Cs clocks \cite{Bauch_03,Bauch_89,Bauch_02}.

For a two-level atom in an intense short near-resonant pulse with
central frequency $\omega$, the Hamiltonian in the interaction
representation with $\ket{\psi}=a_g(t) \exp(-i(\epsilon_g/\hbar +
\omega) t) \ket{g}+a_e(t)\exp(-i\epsilon_e t/\hbar)\ket{e}$, and
rotating wave approximation, takes the form
\begin{equation} \label{Rabi_Ham}
\mathbf{H}=\matwo{\Delta/2 & \Omega^*/2 \\
\Omega/2 & -\Delta/2 } \,,
\end{equation}
where $\Omega= 2\mu A/\hbar$ is the Rabi frequency, $A$ is the slowly
varying envelope of the electric field strength, $\mu$ is the
transition dipole moment, and $\Delta = (\epsilon_e - \epsilon_g -
\hbar\omega)/\hbar$ is the detuning from resonance of the laser
frequency $\omega$.  The solution of the optical Bloch equations for
the two-level atom is given in terms of the unitary evolution operator
for the two-level system for a real slowly varying envelope:
\begin{equation} \label{Rabi_trans}
\mathbf{U}=\matwo{\cos(\frac{\Omega_g t}{2}) -
\frac{i\Delta}{\Omega_g}\sin(\frac{\Omega_g t}{2}) &
\frac{i\Omega}{\Omega_g} \sin(\frac{\Omega_g t}{2}) \\
~&~\\
\frac{i\Omega}{\Omega_g}\sin(\frac{\Omega_g t}{2}) & 
\cos(\frac{\Omega_g t}{2})
+\frac{i\Delta}{\Omega_g}\sin(\frac{\Omega_g t}{2})} \,.
\end{equation}
Here $\Omega_g=\sqrt{|\Omega|^2+\Delta^2}$ is the generalized Rabi
frequency.  In the Ramsey method, the system, assumed to be initially
in the ground state $\ket{g}$, is subjected to two pulses separated by
a delay time $T$,
\begin{equation}
\Omega(t) = \left\{ \begin{array}{ll} 
     \Omega & \mbox{if $0 \leq t \leq \tau_p$}\,, \\
     0 & \mbox{if $\tau_p < t < T+\tau_p$}\,, \\
     \Omega & \mbox{if $T+\tau_p \leq t \leq T+2\tau_p$}\,, 
\end{array} \right.
\end{equation}
with $\Omega \tau_p = \pi/2$ and $T \gg \tau_p$.  From transformation
(\ref{Rabi_trans}), it is clear that the effect of the first pulse is
to evolve the initial ground state $\ket{g}$ into the state
$(\ket{g}+i\ket{e})/\sqrt{2}$.  In a Bloch-sphere picture with
$u=\Re(a_g^*a_e)$, $v=\Im(a_g^*a_e)$, and $w=(|a_e|^2-|a_g|^2)/2$, the
Bloch vector $(u,v,w)$ is projected by the first pulse into the $uv$
plane, as depicted by the red line in Fig.~\ref{fig1}.  During the
delay time, the system carries out phase oscillations, corresponding
to rotation of the Bloch vector in the $uv$ plane with frequency
$\Delta$.  Finally, the second pulse rotates the vector again by an
angle of $\Omega_g\tau_p$ about the $u$ axis, as shown in
Fig.~\ref{fig1}.  Fixing $\Delta$ and measuring the final projection
of the Bloch vector on the $w$ axis as a function of $T$, one obtains
fringes of fixed amplitude $\Omega/\Omega_g$ and frequency $\Delta$.
Alternatively, fixing $T$ and measuring $w(t>T+2\tau_p)$ as a function
of the detuning $\Delta$, results in a power-broadened fringe pattern
of amplitude $\Omega/\Omega_g$ and frequency $2\pi/T$.  The resulting
probability to be in the excited state is given by
\begin{equation}
P_e = \frac{4\Omega^2}{\Omega_g^2}\sin^2(\frac{1}{2} \Omega_g \tau_p)
\left(-\cos(\frac{1}{2}\Omega_g\tau_p) \cos(\frac{1}{2}\Delta T) +
\frac{\Delta}{\Omega_g}\sin(\frac{1}{2}\Omega_g\tau_p)
\sin(\frac{1}{2}\Delta T)\right)^2 \,.
\end{equation}

\begin{figure}
\centering
\includegraphics[scale=0.9]{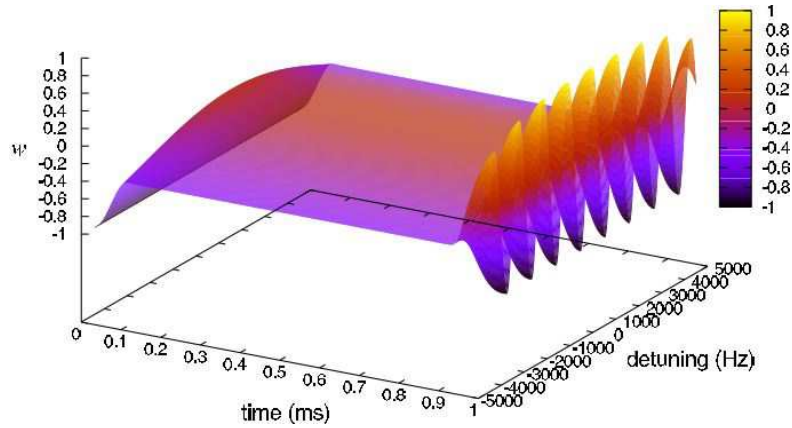}
\caption{(color online) Population inversion $w$ versus time and
detuning $\Delta$ in a Ramsey separated field method.  The interaction
strength is taken to be $\kappa = 100$ Hz.  \\}
\label{fig2}
\end{figure} 

Figure~\ref{fig2} shows the population inversion $w$ versus time and
detuning $\Delta$ using a Ramsey separated field method for one atom
in an optical lattice site.  The final time corresponds to the time at
which the Ramsey clock signal is measured as a function of detuning
$\Delta$, i.e., either the population of the ground or the excited
state is measured as a function of $\Delta$.  Note that the excited
state population at the final time is unity at zero detuning and that
the population inversion oscillates as a function of detuning.

It is easy to generalize this treatment to a time-dependent Rabi
frequency $\Omega(t) = 2\mu A(t)/\hbar$ due to a pulse of light which
turns on and off with a finite rate in terms of the pulse area,
$\int_0^{\tau_p} \Omega(t')dt'$.  For a $\pi/2$ pulse,
$\int_0^{\tau_p} \Omega(t')dt' = \pi/2$, where $\tau_p$ is the pulse
duration.  For the Ramsey pulses, $\int_{0}^{\tau_p} \Omega(t')dt' =
\pi/2$, $\int_{\tau_p}^{T+\tau_p} \Omega(t')dt' = 0$, and
$\int_{T+\tau_p}^{T+2\tau_p} \Omega(t')dt' = \pi/2$.

\section{Second-quantized Ramsey model} \label{section3}

In order to study the effects of collisions on Ramsey fringes obtained
in a separated-fields scheme, we use a second-quantized
formalism to treat a multiply-populated single site of the 
optical potential and calculate the Ramsey fringe dynamics.  The
many-body Hamiltonian for a system of two-level atoms, all in the same
external motional state of a trap, can be written as \cite{Vardi01,Anglin_01}
\begin{equation}
\label{hf}
\hH = \sum_{i=g,e} E_{i} \ha^\dag_{i}\ha_{i} - \frac{\hbar
\Omega(t)}{2} (\ha^\dag_{g} \ha_{e} + \ha^\dag_{e} \ha_{g}) +
\sum_{i,j=g,e} G_{ij} \ha^\dag_{i} \ha^\dag_{j}\ha_{j}\ha_{i} \,.
\end{equation}
Here the subscripts $g$ and $e$ indicate the ground and excited states
of the two level atom, and $\ha_i$ is the annihilation operator for an
atom in internal state $i$, where $i=g,e$.  The self- and cross-
atom-atom interaction strengths are denoted as $G_{ii}$ and $G_{ij}$
respectively, the internal energy of state $i$ is denoted as $E_i$,
where $E_g = \epsilon_g + \hbar \omega$ and $E_e = \epsilon_e$.  For
bosonic atoms, the creation and annihilation operators satisfy the
commutation relations
\begin{equation}
[\ha_i,\ha_j^\dag] = \delta_{ij} \,, \qquad [\ha_i^\dag,\ha_j^\dag] =
0 \,, \qquad [\ha_i,\ha_j] = 0 \,.
\end{equation}
Defining the operators
\begin{equation}
\hn_i = \ha^\dag_{i} \ha_{i} , \; \; \;
\hN = \hn_g + \hn_e , \; \; \;
\hL_z = {\hn_g - \hn_e \over 2}  , \; \; \;
\hL_x = {\ha^\dag_g \ha_e + \ha^\dag_e \ha_g \over 2}  , \; \; \;
\hL_y = {\ha^\dag_g \ha_e - \ha^\dag_e \ha_g \over 2i} ,
\end{equation}
the Hamiltonian can be written as
\begin{equation}
\label{Ham}
\hH = E \hN + \hbar \Delta \hL_z - \hbar \Omega(t) \hL_x +
\sum_{i=1}^{2} G_{ii} (\hn_{i}^2 - \hn_{i}) + 2 G_{ge} \hn_{g} \hn_{e}
\,,
\end{equation}
where
\begin{eqnarray}
E &=& (\hbar \omega + \epsilon_g + \epsilon_e)/2 \,, \\
\hbar \Delta &=& (\hbar \omega - \epsilon_e + \epsilon_g) \,.
\end{eqnarray}
Since $\hN$ commutes with $\hH$, the Hamiltonian conserves the total
number of particles.  Typically, in a Ramsey-fringe experiment, the
initial state is assumed to have all $N$ atoms in the ground state
$\ket{g}$.  Fixing $\hN=N$ in the single site that we are considering
here (i.e., so that there are no fluctuations), and using the
identities $\hn_g = \hN/2 + \hL_z$ and $\hn_e = \hN/2 - \hL_z$, we
finally obtain
\begin{eqnarray}
\label{Hamilt}
\hH&=&(E - {G_{gg} + G_{ee} \over 2}) \hN + (G_{gg} + G_{ee} +
2G_{ge}) {\hN^2 \over 4}\nonumber\\ 
&&- \hbar \Omega(t) \hL_x+[\hbar \Delta+(G_{gg} - G_{ee}) (\hN -1)] \hL_z  
\nonumber\\
&&+(G_{gg} + G_{ee} - 2G_{ge}) \hL_z^2\, .
\end{eqnarray}
Since Ramsey spectroscopy measures essentially single-particle
coherence, we will be interested in the dynamics of the reduced
single-particle density matrix (SPDM) $\rho(t)$:
\begin{equation}
\rho(t)=\langle\hrho(t)\rangle=
\frac{N}{2}{\cal I}+u(t)\sigma_x+v(t)\sigma_y+w(t)\sigma_z\, ,
\end{equation}
where $\hrho(t)=\ha_i^\dag(t)\ha_j(t)$ is the reduced single-particle
density operator, ${\cal I}$ is a two-by-two unity matrix,
$\sigma_i,i=x,y,z$ are Pauli matrices, and $u=\langle \hL_x\rangle$,
$v=\langle \hL_y\rangle$, $w=\langle \hL_z\rangle$ are the components
of the single-particle Bloch vector, corresponding to the real- and
imaginary parts of the single-particle coherence, and to the
population imbalance between the two modes, respectively.  The
expectation values, $\langle \cdot \rangle$ are over the $N$-particle
states.  The Liouville-von-Neuman equation for the SPDM,
\begin{equation}
\label{lvn}
{d \over dt} \rho = {i \over \hbar} \langle[\hrho,\hH]\rangle ={i
\over \hbar} \sum_{j=x,y,z}\langle[\hL_j,\hH]\rangle \, \sigma_j~,
\end{equation}
is thus equivalent to the expectation values of the three coupled
Heisenberg equations of motion for the $SU(2)$ generators,
\begin{eqnarray}
\label{eomlx}
{d \over dt} \hL_x &=&-[\Delta+\gl(\hN -1)]\hL_y
-\gnl(\hL_y\hL_z+\hL_z\hL_y), \\
\label{eomly}
{d \over dt} \hL_y &=&[\Delta+\gl(\hN -1)]\hL_x
+ \gnl (\hL_x\hL_z + \hL_z \hL_x)\nonumber\\
&&+\Omega(t)\hL_z~, \\
\label{eomlz}
{d \over dt} \hL_z &=& -\Omega(t) \hL_y \,,
\end{eqnarray}
where we denote the linear and nonlinear interaction strength 
parameters respectively by
\begin{equation} 
\gl\equiv(G_{gg} - G_{ee})/\hbar~,
\end{equation}
\begin{equation} \label{GNL}
\gnl\equiv (G_{gg} + G_{ee} - 2G_{ge})/\hbar~.
\end{equation}

In order to study effects resulting in from the coupling to an
environment consisting of a bath of external degrees of freedom, we
use the master equation \cite{Carmichael_93,Gardiner_99}
\begin{equation}
\label{rho}
{d \over dt} \hrho = {i \over \hbar} [\hrho,\hH] - \sum_{k} \Gamma_k 
(2\hO_k \hrho\hO_k^{\dag} - \{\hO_k^{\dag} \hO_k,\hrho\}) \,,
\end{equation}
where the second term on the right hand side of (\ref{rho}) has the
Markovian Lindblad form \cite{Lindblad} and gives rise to dissipation
effects of the bath on the density operator.  Such terms may be used
to depict decay due to spontaneous emission, atom-surface
interactions, motional effects, black-body radiation and other
environmental effects.  The Lindblad operators $\hO_k$ are determined
from the nature of the system-bath coupling and the coefficients
$\Gamma_k$ are the corresponding coupling parameters.  In what follows
we shall assume that the most significant dissipation process is
dephasing, and we neglect all other dissipation effects (e.g., we
assume spontaneous emission is negligible because the lifetime of the
excited state is much longer than the Ramsey process run-time, etc.).
Thus, the Lindblad operator is taken to be ${\hat L}_z$, yielding
dephasing of the transition dipole moment without affecting the
population of the ground or excited states.  We shall only study this
type of $t_2$ dephasing, without fully exploring the effects of other
types of Lindblad operators.  Clearly, it is easy to generalize this
and consider the effects of additional Lindblad operators, but we do
not do so here.  Moreover, non-Markov treatments can be used to
generalize the Markovian approximation made in deriving
Eq.~(\ref{rho}) \cite{Carmichael_93,Gardiner_99}.

\begin{figure}
\centering
\includegraphics[scale=0.8]{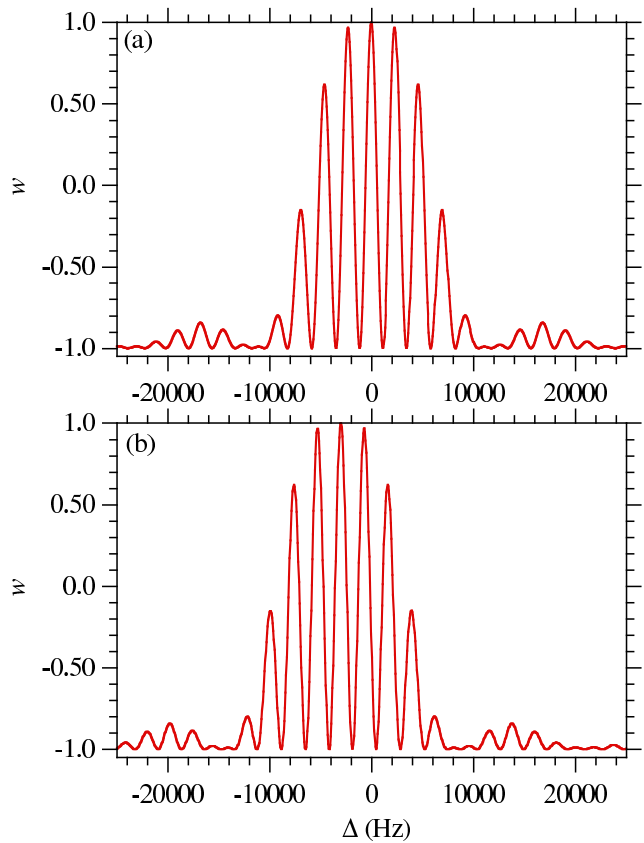}
\caption{(color online) Linear collisional shift.  Ramsey fringes for
$N$ particles in a lattice site, for (a) $\gl=\gnl=0$ and (b)
$\gl(N-1)=3000$ Hz, $\gnl=0$.}
\label{fig3}
\end{figure}

\section{Single-site many-body dynamics}\label{section4} 

We first study how interparticle interactions can modify the
Bloch-vector dynamics in a single-site Ramsey scheme.  From
Eq.~(\ref{Hamilt}) and Eqs.~(\ref{eomlx})-(\ref{eomlz}) it is evident
that there are three possible collisional effects.  First, differences
between the interaction strengths of ground and excited atoms, will
induce a linear frequency shift of the center of the Ramsey fringe
pattern by an amount $\gl(N-1)$.  This effect is illustrated in
Fig.~\ref{fig1}, where the trajectory of the Bloch vector during a
Ramsey sequence is traced in the absence of interactions (red) and in
their presence (blue).  Clearly, the frequency of oscillation in the
$uv$ plane during the delay time $T$ is modified by the interaction.
In Fig.~\ref{fig3} we plot the resulting fringe pattern when $\gnl$ is
set equal to zero.  Fig.~\ref{fig3}(a) depicts the Ramsey fringes
without any interaction, $\gl=\gnl=0$, whereas in Fig.~\ref{fig3}(b)
we plot the fringe pattern with $\gl=3000$ Hz.  Since for $\gnl=0$
Eqs.~(\ref{eomlx})-(\ref{eomlz}) remain linear, and the only effect is
a linear frequency shift.  The whole curve $w(\Delta)$ versus $\Delta$
is simply shifted in $\Delta$ by an amount $\gl(N -1)$.

Also shown in Fig.~\ref{fig1} is the loss of single-particle
coherence due to interparticle entanglement \cite{Vardi01}.  This
dephasing process is only possible for the nonlinear case when
$\gnl \neq 0$.  It is manifested in the reduction of the
single-particle purity $p = \mathrm{Tr}(\rho^2)=|u|^2+|v|^2+|w|^2$,
during the evolution.  In comparison, for the single-particle case
depicted in red, single-particle purity is trivially conserved and the
length of the Bloch vector is unity throughout the propagation time; 
the Bloch vector at the end of the process sits well within the Bloch 
sphere.

\begin{figure}
\centering
\includegraphics[scale=0.8]{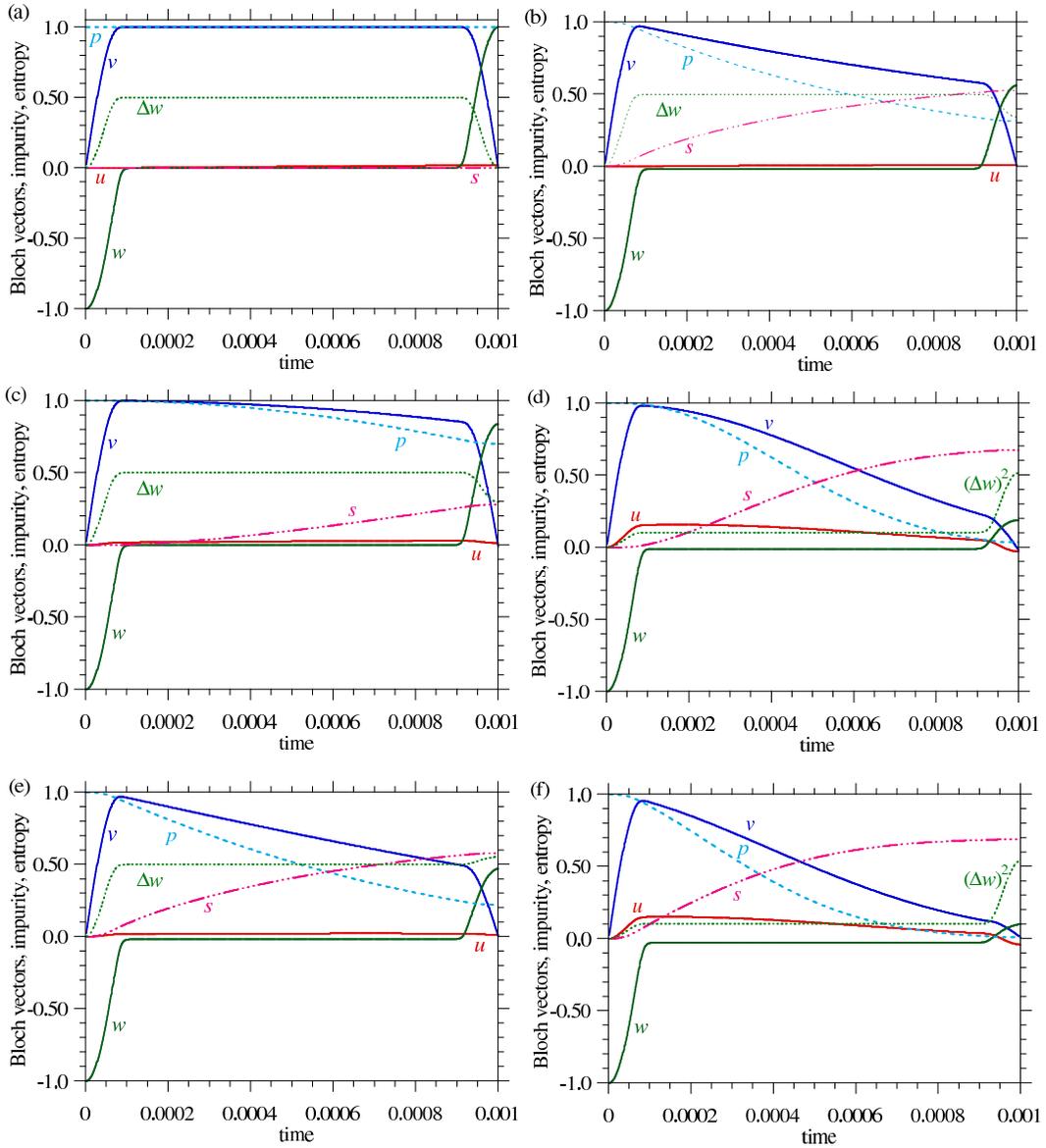}
\caption{(color online) Bloch vector, $u$, $v$, $w$, purity $p$,
entropy $s$ and variance of $w$ versus time using a Ramsey separated
field method for zero detuning $\Delta$.  (a) Two atoms without
dephasing and without interaction.  (b) Two atoms with dephasing
($\Gamma_{L_z} = 100$ Hz) and without interaction.  (c) Two atoms
without dephasing but with interaction ($\kappa = 100$ Hz).  (d) Ten
atoms without dephasing but with interaction ($\kappa = 100$ Hz).  (e)
Two atoms with dephasing ($\Gamma_{L_z} = 100$ Hz) and interaction
($\kappa = 100$ Hz).  (e) Ten atoms with dephasing and interaction.
\\}
\label{fig4}
\end{figure}

The decay of single particle coherence due to entanglement is
illustrated in Fig.~\ref{fig4}, where the Bloch vector ($u$, $v$,
$w$,), the single particle purity $p$, the single-particle entropy $s
= \mathrm{Tr}(\rho\ln\rho) =-\,\ln[(1+\sqrt{p})^{1+\sqrt{p}}
(1-\sqrt{p})^{1-\sqrt{p}}/4]/2$, and the variance of $w$, $\Delta w$,
are plotted versus time in a Ramsey separated field method for $\Delta
= 0$.  The total time for the Ramsey process is taken to be
$t_{\mathrm{tot}} = T+2\tau_p = 1.0 \times 10^{-3}$ s, with the first
and second Ramsey pulses each of duration $t_{\mathrm{tot}}/12$, and
the Rabi frequency of the pulses are $3.0\times 10^{3}$ Hz, so the two
pulses each have pulse area $\pi/2$.  The interaction strengths were
arbitrarily set to $G_{ee} = G_{ge} = 0$, so that $\gl = \gnl =
G_{gg}/\hbar\equiv\kappa$.  In frame (a) $\kappa$ is set to zero
(corresponding to the case of one atom per site in the optical
lattice).  The effect of decoherence due to the $\Gamma_{L_z}$ term in
Eq.~(\ref{lvn}) is shown in frame (b), while collisional dephasing is
depicted in frames (c)-(e) were we set $\kappa=100$ Hz.  In frames (c)
and (e) we have taken two atoms per site, whereas in frames (d) and
(f) we have taken 10 atoms per site.  In frames (e) and (f), dephasing
with strength $\gamma_z = 1.0 \times 10^{2}$ Hz is included.  From
comparison of Fig.~\ref{fig4}(b) and Fig.~\ref{fig4}(c-d), it is clear
that the loss of single-particle coherence due to entanglement is
similar to the effect of dephasing due to coupling to an external
bath.  In particular, the final population inversion is strongly
affected.  Comparing Figs.~\ref{fig4}(b) and Fig.~\ref{fig4}(c), it is
clear that collisional dephasing is stronger as the number of
particles grows.  The combined effect of decoherence and 
interparticle entanglement is shown in frames (e) and (f).

\begin{figure}
\centering
\includegraphics[angle=-90,scale=0.6]{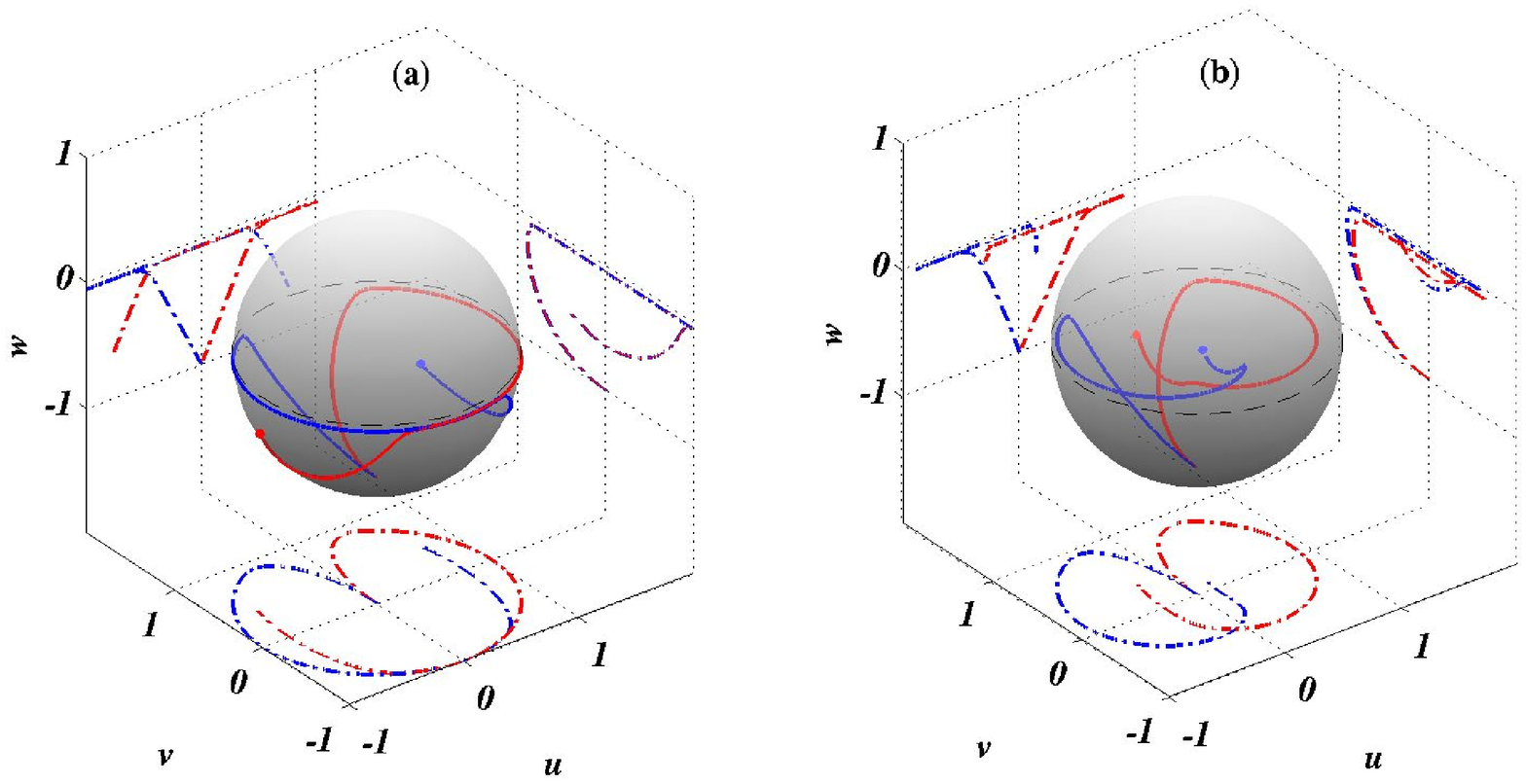}
\caption{(color online) Collisional asymmetry of Ramsey fringes.  Red
and blue curves correspond to Bloch-vector trajectories with positive
and negative detuning, respectively.  The final position of the Bloch
vectors is indicated by a dot at the end of the trajectories.  In the
absence of interactions (a) the final projection onto the $w$ axis is
independent of the sign of $\Delta$.  However, when interactions are
present (b) nonlinear collisional shifts break the symmetry.
Dash-dotted lines depict projections of the trajectory onto the $uv$,
$uw$, and $vw$ planes.}
\label{fig5}
\end{figure}

\begin{figure}
\centering
\includegraphics[angle=0,scale=0.6]{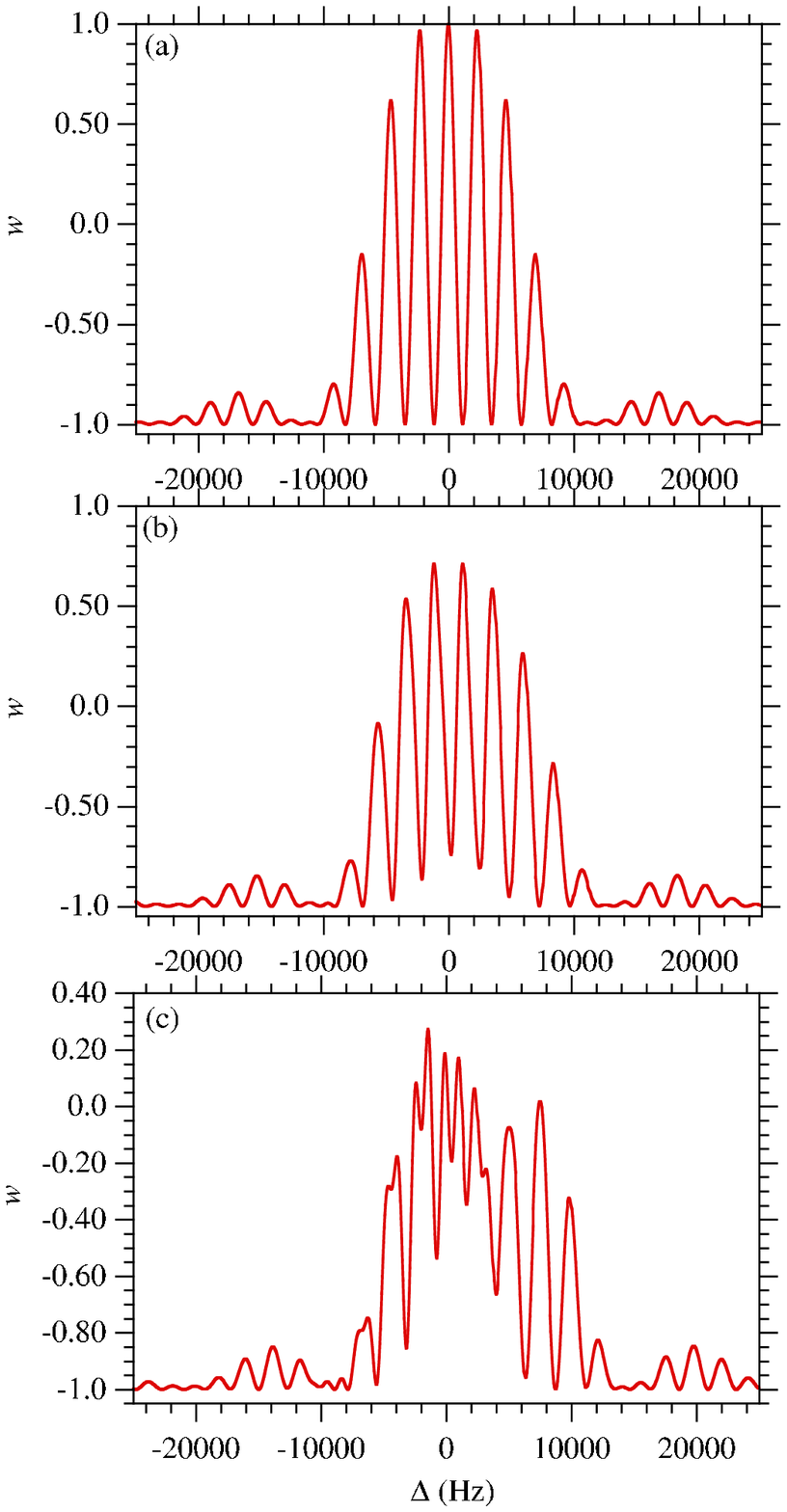}
\caption{(color online) Asymmetric Ramsey fringes.  Numerically
calculated Ramsey fringe patterns, at three different values of the
nonlinear interaction: (a) $\gnl=0$, $\gl=0$, (b) $\gnl=1500Hz$,
$\gl=0$, and (c) $\gnl=3000$ Hz, $\gl=0$.}
\label{fig6}
\end{figure}

Yet another collisional effect is caused by the $\gnl\hL_z^2$ term in
the Hamiltonian (\ref{Hamilt}).  In the mean-field approximation,
replacing $\hL_i$ by their expectation values, this term leads to a
nonlinear frequency shift of $2\gnl w(\tau_p)$, where $w(\tau_p)$ is
the value of the population imbalance after the first pulse.  As
illustrated in Figs.~\ref{fig5} and \ref{fig6}, the nonlinear shift
results in an asymmetric fringe pattern, because for $\gnl\neq 0$,
$w(\tau_p)$ depends on the sign of $\Delta$ during the first pulse.
Without interactions, when $\gl=\gnl=0$, the dynamical equations
(\ref{eomlx})-(\ref{eomlz}) are symmetric under $\Delta\rightarrow
-\Delta~,~\hL_x\rightarrow -\hL_x$.  This symmetry shows up in
Fig.~\ref{fig5}(a) where the evolution with two opposite values of
$\Delta$ is traced over the Bloch sphere.  As shown in the various
projections onto the $uv$, $vw$, and $uw$ planes,
$u_{\Delta}(t)=-u_{-\Delta}(t)$, $v_{\Delta}(t)=v_{-\Delta}(t)$, and
$w_{\Delta}(t)=w_{-\Delta}(t)$, where
$(u_\Delta(t),v_\Delta(t),w_\Delta(t))$ denote the Bloch vector
$(u(t),v(t),w(t))$, evolved in a Ramsey sequence with detuning
$\Delta$.  Similarly, when $\gnl=0$ and $\gl$ is nonvanishing, the
resulting shifted pattern is still symmetric about $\Delta=-(N-1)\gl$
since the dynamical equations are invariant under $\Delta +
(N-1)\gl\rightarrow -(\Delta+(N-1)\gl)~,~\hL_x \rightarrow -\hL_x$.
However, for finite values of $\gnl$ the symmetry breaks down because
the last terms on the r.h.s. of Eqs.  (\ref{eomlx}) and (\ref{eomly})
change sign, as demonstrated in Fig.~\ref{fig5}(b).  Consequently,
when the final projection on the $w$ axis is plotted as a function of
the detuning $\Delta$, we obtain asymmetric fringe patterns.  In
Fig.~\ref{fig6}, we compare the symmetric single-particle pattern
(Fig.~\ref{fig6}a) with asymmetric two-particle Ramsey fringes
obtained with nonvanishing values of $\gnl$ (Figs.~\ref{fig6}b and
\ref{fig6}c).  The linear shift $\gl$ is set equal to zero for all
plots.  Both the asymmetry of the emerging lineshape and the reduction
in fringe contrast due to collisional dephasing are evident.

\begin{figure}
\centering
\includegraphics[scale=0.8]{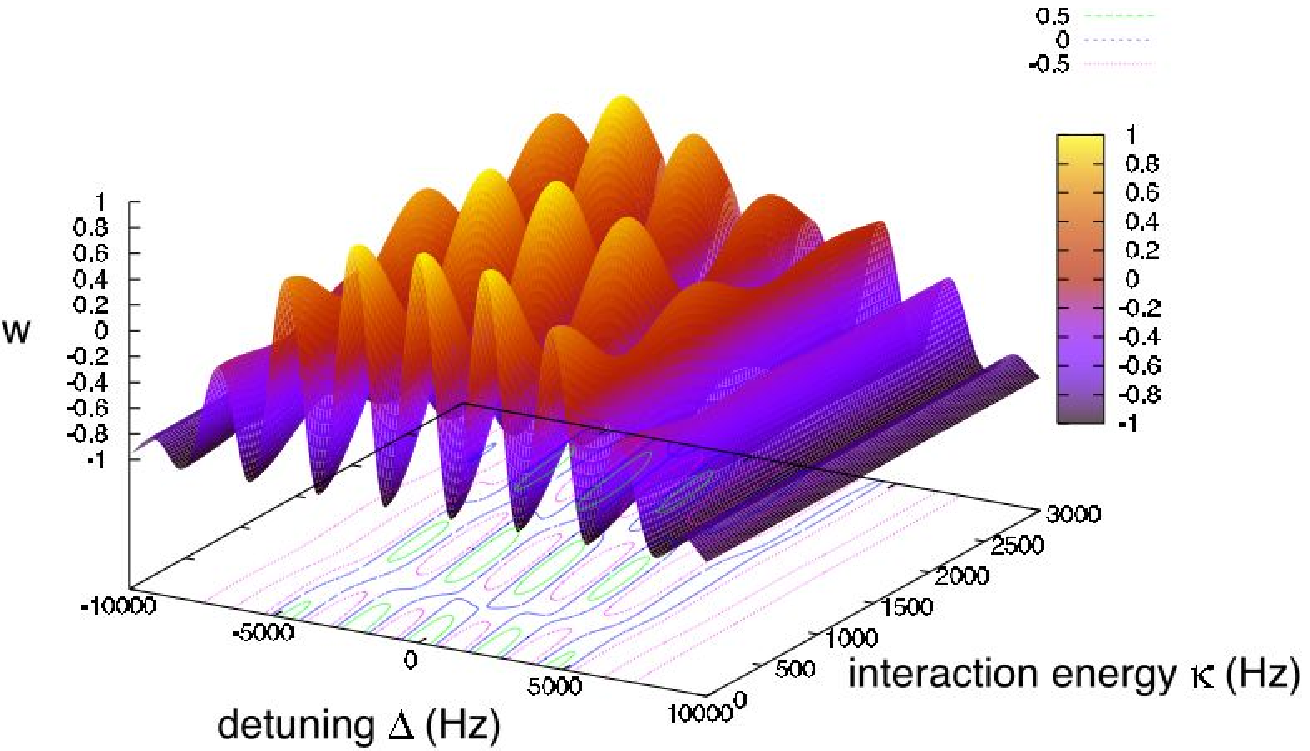}
\caption{(color online) Population inversion $w$, versus detuning
$\Delta$ and interaction strength $\gl=\gnl=\kappa$ in a Ramsey
separated field configuration at the final time after the second
pulse.  \\}
\label{fig7}
\end{figure}

The combined effect of the linear shift and the nonlinear term
in Eqs.~(\ref{eomlx})-(\ref{eomlz}) is shown in Fig.~\ref{fig7}
where the final population inversion $w(t=\tau+2\tau_p)$ is plotted
as a function of the detuning $\Delta$ and the interaction strength,
set arbitrarily to $\gl=\gnl=\kappa$. Fringe maxima are located about 
$(\Delta+\kappa (N-1)+2\kappa w(\tau_p))=2j\pi$ where $j$ is an integer
and $w(\tau_p)$ depends on both $\Delta$ and $\kappa$. As $\kappa$
increases, collisions shift the peak intensity and the fringe 
contrast is reduced due to collisional dephasing.

\section{Two-site Many-Body Dynamics}  \label{section5}

Having established how interparticle interactions affect the
Ramsey lineshapes in a multiply-occupied single lattice
site, we proceed to consider an optical lattice with multiple 
sites labeled by the site index $\alpha$.  Each site is 
populated with atoms that can be in any one of two levels, 
the ground state level $|g\rangle$ and the excited state level 
$|e\rangle$, coupled through a Rabi flopping term in the
Hamiltonian.  The full Hamiltonian is given by
\begin{subequations}
  \label{H-orig}
\begin{eqnarray}
\hH &=& \sum_{\langle \alpha,\beta \rangle}
\hH_{\mathrm{hop}}^{(\alpha,\beta)} \\
&&+ \sum_\alpha \left[
\hH_{\mathrm{int}}^{(\alpha)} + \hH_{\mathrm{diff}}^{(\alpha)} +
\hH_{\mathrm{sum}}^{(\alpha)} + \hH_{\mathrm{R}}^{(\alpha)}\right]\ , 
\nonumber\\
\hH_{\mathrm{hop}}^{(\alpha,\beta)} &=& \hbar
J_{\mathrm{hop}}[(\had_{\alpha} \ha_\beta + h.c.) +
(\hbd_{\alpha}\hb_\beta + h.c.)] \ , \\
\hH_{\mathrm{int}}^{(\alpha)} &=& \tfrac{1}{2}U_{aa}\had_\alpha
\had_\alpha\ha_\alpha\ha_\alpha + \tfrac{1}{2}U_{bb} \hbd_\alpha
\hbd_\alpha\hb_\alpha\hb_\alpha\nonumber\\
&&+U_{ab}\had_\alpha \hbd_\alpha \hb_\alpha\ha_\alpha \ , \\
\hH_{\mathrm{diff}}^{(\alpha)} &=& \frac{\hbar\Delta}{2} (\had_\alpha
\ha_\alpha - \hbd_\alpha\hb_\alpha)\ , \label{H-mag} \\
\hH_{\mathrm{sum}}^{(\alpha)} &=& \frac{\epsilon_a + \hbar \omega +
\epsilon_b}{2} (\had_\alpha\ha_\alpha + \hbd_\alpha\hb_\alpha)\ ,
\label{H-el} \\
\hH_{\mathrm{R}}^{(\alpha)} &=& \frac{\hbar\Omega(t)}{2}(\had_\alpha
\hb_\alpha e^{i\phi_\alpha(t)} + \hbd_\alpha \ha_\alpha
e^{-i\phi_\alpha(t)})\ .  \label{H-las}
\end{eqnarray}
\end{subequations}
Here, the operators $a$ and $b$ are bosonic destruction operators for
atoms in the two states $|g\rangle$ and $|e\rangle$, and the indices
$\alpha$ and $\beta$ run over the lattice sites, with $\langle
\alpha,\beta \rangle$ denoting a pair of adjacent sites.  The
constants $J_{\mathrm{hop}}$ and $U_{ij}$, $i,j = g,e$, are the
strength of the hopping to neighboring sites in $H_{\mathrm{hop}}$ and
the effective on-site interaction in $H_{\mathrm{int}}$, respectively.
$H_{\mathrm{diff}}$ and $H_{\mathrm{sum}}$ are the energy difference
and average of the dressed states of the atoms, and $H_{R}$ induces
Rabi transitions between the two atomic internal states with Rabi
frequency $\Omega(t)$ and phase $\phi_\alpha(t)$, which are related to
the intensity and phase of the laser which induces these transitions.
The Hamiltonian in (\ref{H-orig}) is identical to what we used in
previous sections, except for the addition of the hopping term
$\hH_{\mathrm{hop}}$ that can result in hopping of particles to
adjacent sites with rate $J_{\mathrm{hop}}$.

\begin{figure} \centering
\includegraphics[scale=0.9]{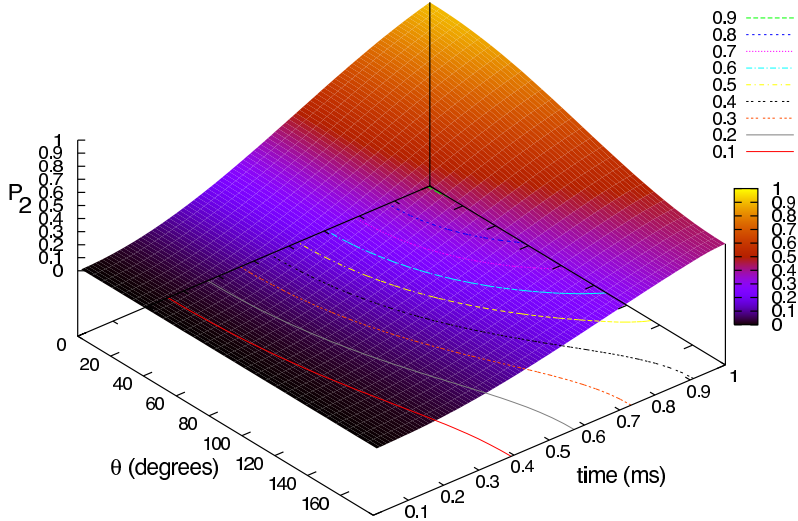}
\caption{(color online) Probability for double occupancy $P_2$, during
a Ramsey sequence with $\Delta=0$, versus relative Rabi-drive phase
angle $\theta$ for $J_{\mathrm{hop}} = 100$ Hz.  Tunneling is
suppressed for $\theta=\pi$ due to destructive interference and
enhanced for $\theta=0$ due to constructive interference.
Interparticle interactions are set equal to zero.  \\}
\label{fig8} \end{figure}

Note that the phase $\phi_\alpha(t)$ can depend upon the site index
$\alpha$.  Consider, for simplicity, a 1D optical lattice along the
$x$ axis, and a plane wave field with detuning $\Delta$ from resonance
with wave vector ${\mathbf k}$.  The Rabi frequency at site $\alpha$
is given by $\Omega({\mathbf r}_\alpha,t) = \Omega(t) \exp[i{\mathbf
k} \cdot {\mathbf r}_\alpha] = \Omega(t) \exp(ik_x x_\alpha)$.  Thus,
there is a phase difference $\theta$ between the Rabi frequency at
neighboring sites, $\theta = k_x \delta x$, where $\delta x$ is the
lattice spacing.  The angle between the wave vector ${\mathbf k}$ and
the $x$ axis can be adjusted to control the phase difference
$\theta=\phi_\alpha - \phi_{\alpha-1}$.  In what follows, we show that
a proper choice of the relative phase angle $\theta$ between Rabi
drive fields in adjacent sites may be used to suppress the tunneling
between them and thus reduce collisional shifts.  When
$J_{\mathrm{hop}} = 0$, there is no relevance to the phases
$\phi_\alpha(t)$ (they do not affect the dyanmics), but as soon as
hopping from one site to an adjacent site is allowed, these phases can
affect the dynamics.

\begin{figure} \centering
\includegraphics[scale=0.7]{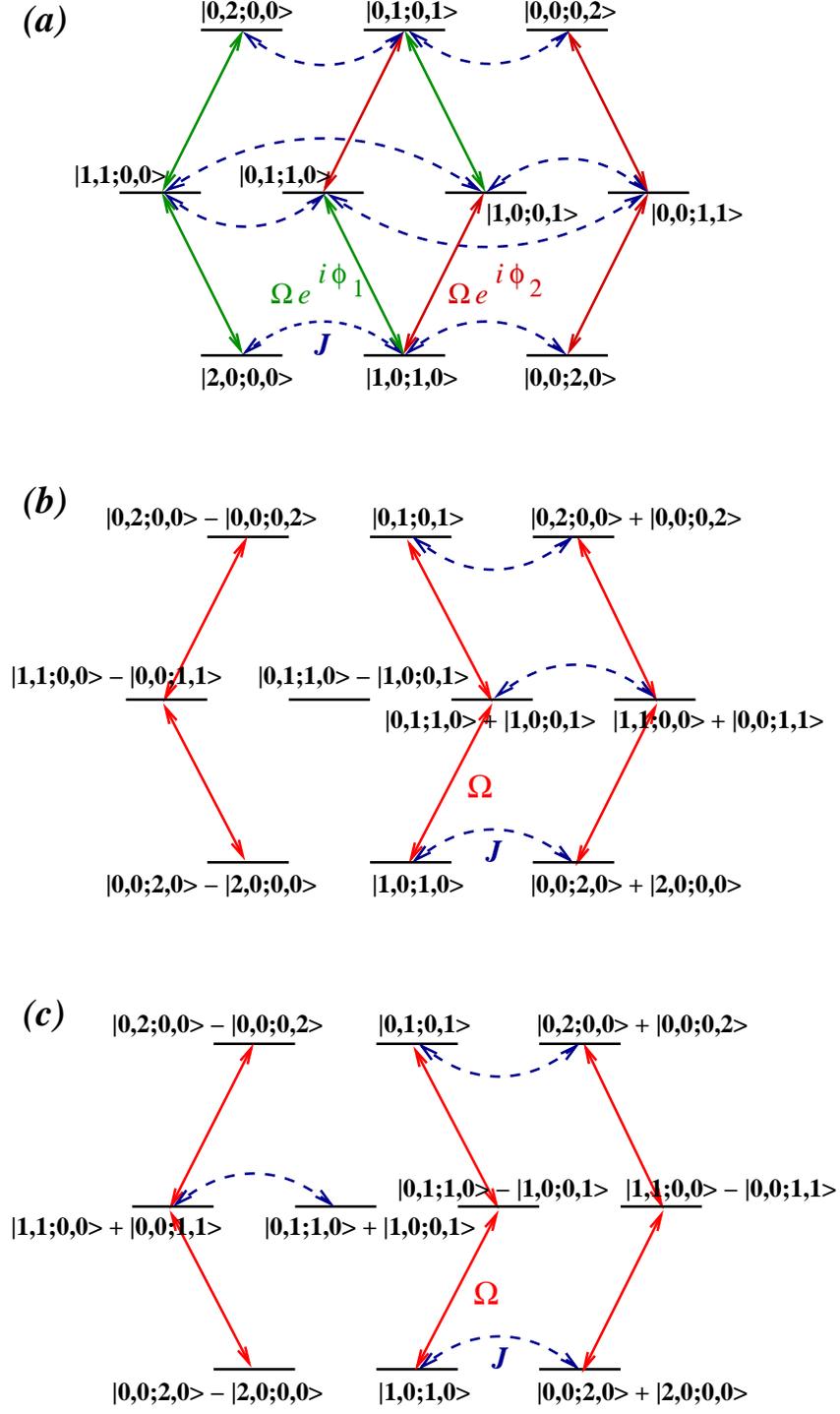}
\caption{(color online) Two-particle level schemes for Ramsey
spectroscopy in the presence of tunneling: (a) general scheme (b)
coupling between parity eigenstates for $\theta=0$ (c) coupling
between parity eigenstates for $\theta=\pi$.  Rabi coupling between
ground- and excited states, is denoted by red and green solid arrows,
corresponding to different phases of the driving fields in adjacent
sites.  Blue dashed arrows denote hopping between sites.  Fock states
are denoted as $\ket{n^g_1,n^e_1;n^g_2,n^e_2}$, wh ere
$n^g_\alpha=\langle\ha_\alpha^\dag\ha_\alpha\rangle$ and
$n^e_\alpha=\langle\hb_\alpha^\dag\hb_\alpha\rangle$.  \\}
\label{fig9} 
\end{figure}

We expect that for sufficiently low densities, the probability of
finding more than two adjacent populated sites is very small.  It is
therefore possible to capture much of the physics of the hopping
process, using a two-site model.  At higher densities, more elaborate
models should be used.  In Fig.~\ref{fig8} we plot numerical results
for two particles in two sites ($N=2,\alpha=1,2$) for
$J_{\mathrm{hop}} = 100$ Hz, showing the probability of double
occupancy during a Ramsey fringe sequence in the presence of
tunneling, as a function of the relative angle $\theta$.  The
parameters used are the same as those used previously, except that now
we allow for hopping, and therefore $J_{\mathrm{hop}}$ and the angle
$\theta$ affects the dynamics.  The initial conditions correspond to a
single atom in its ground state in each lattice site before the Ramsey
process begins.  When both atoms are driven in-phase ($\theta=0$),
tunneling takes place, leading to multiple occupancy.  The tunneling
is significantly suppressed for a $\theta=\pi$ phase difference
between the Rabi drives in adjacent sites.  As shown in
Fig.~\ref{fig9}, this suppression originates from destructive
interference.  In Fig.~\ref{fig9}a we plot the level scheme for a
two-particle, two-site system.  The levels
$\ket{n^g_1,n^e_1;n^g_2,n^e_2}$ denote Fock states with $n^g_\alpha$
and $n^e_\alpha$ particles in the ground- and excited states
respectively, of site $\alpha$.  Depending on the relative phase
between the optical drive fields, tunneling from states with a ground
state atom in one site and an excited atom in another, can interfere
constructively or destructively.  Therefore, as seen in
Fig.~\ref{fig9}b and Fig.~\ref{fig9}c, tunneling can only take place
between even-parity states.  For in-phase Rabi drive $\theta=0$
(Fig.~\ref{fig9}b), the initial even-parity state $\ket{1,0;1,0}$ is
coupled by a single optical excitation to the even-parity state
$(1/\sqrt{2})\left(\ket{0,1;1,0}+\ket{1,0;0,1}\right)$, for which the
two tunneling pathways interfere constructively.  In contrast, for
out-of-phase drive $\theta=\pi$ (Fig.~\ref{fig9}c), the first $\pi/2$
pulse drives the system partially into the odd-parity state
$(1/\sqrt{2})\left(\ket{0,1;1,0}-\ket{1,0;0,1}\right)$ which is 'dark'
to tunneling, leading to a reduced probability of multiple occupancy
during the time evolution.

\begin{figure} \centering
\includegraphics[scale=0.7]{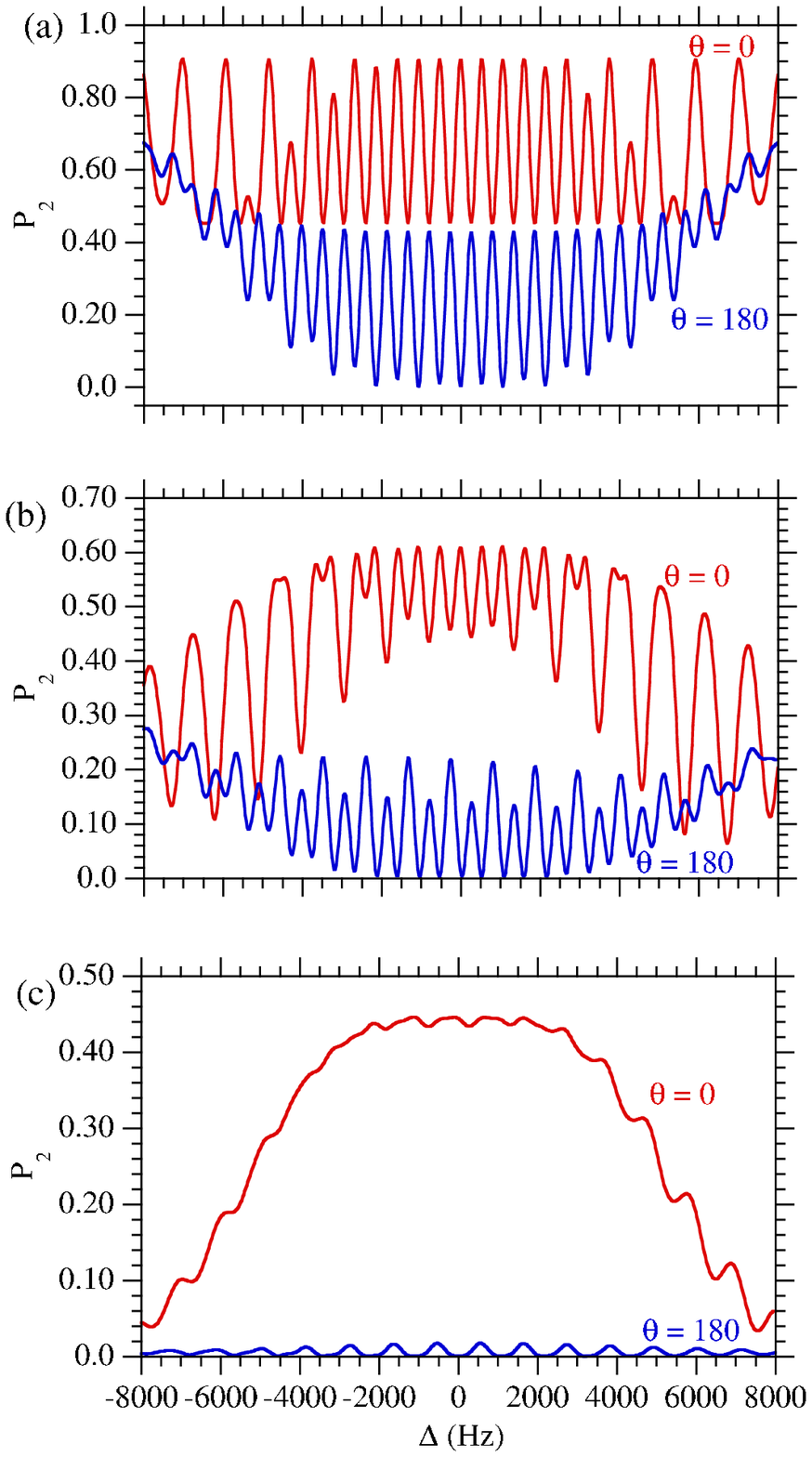}
\caption{(color online) Probability of double-occupation at the end of
a Ramsey sequence versus detuning, for different values of the
nonlinearity $\gl=\gnl=\kappa$: (a) $\kappa=0$, (b) $\kappa =
5.0\times 10^{2}$, and (c) $\kappa = 1.0\times 10^{3}$.  Red curves
correspond to $\theta=0$ whereas blue curves show the lineshape for
$\theta=\pi$.  \\} \label{fig10} \end{figure}

\begin{figure} \centering
\includegraphics[scale=0.7]{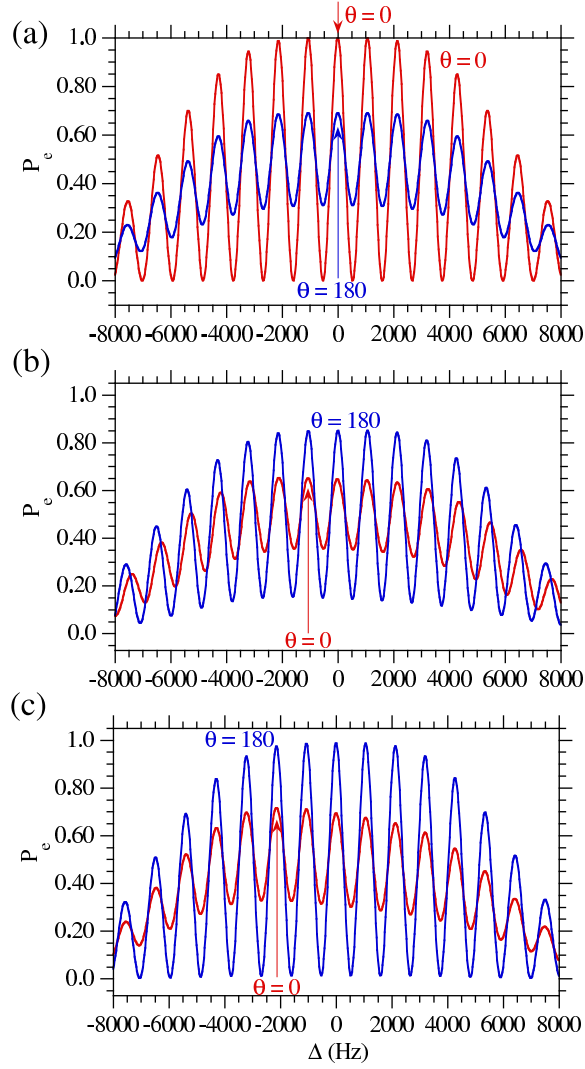}
\caption{(color online) Ramsey fringe patterns for different values of
the nonlinearity $\gl=\gnl=\kappa$: (a)$\kappa=0$, (b) $\kappa =
5.0\times 10^{2}$, and (c) $\kappa = 1.0\times 10^{3}$.  Red curves
correspond to $\theta=0$ whereas blue curves show the lineshape for
$\theta=\pi$.  The arrows serve to roughly indicate the centers of the
distributions.  \\} \label{fig11} \end{figure}

The lower probability of finding multiply occupied sites in an
inverted-phase Rabi-drive configuration, is manifested in the emerging
Ramsey fringe patterns.  In Fig.~\ref{fig10} we plot the probability
of double-occupancy at the end of a Ramsey sequence, as a function of
detuning, for different degrees of nonlinearity which is arbitrarily
set to $\gl = \gnl = \kappa$.  While increasing nonlinearity generally
tends to localize population and reduce tunneling, the probability of
finding both particles in the same lattice site is always lower for
out-of-phase driving.  The resulting Ramsey lineshapes are plotted in
Fig.~\ref{fig11}, clearly demonstrating reduced collisional effects
for $\theta=\pi$ (blue curves), compared to the in-phase driving fields
scheme (red curves).  The center of the distributions for $\theta=0$ 
are clearly strongly shifted from $\Delta = 0$, and much less shifted 
for $\theta=\pi$.  This reduced shift in the center of the 
distribution is, of course, of central importance for a clock.

\section{Conclusion} \label{conclusion}
Optical lattice atom clocks hold great promise for setting frequency
standards.  In order to achieve high accuracy with boson atoms,
collisional frequency shifts have to be taken in account.  We have
shown that collisions can degrade Ramsey lineshapes by shifting their
centers, rendering them asymmetric, and by reducing fringe-visibility
due to the loss of single-particle coherence.  Considering two
adjacent populated sites in a 1D optical lattice configuration, we
propose a method of reducing dynamical multiple population of lattice
sites, based on driving different sites with different phases of the
Rabi drive.  Due to destructive interference between tunneling
pathways leading to states with one ground and one excited atom in the
same site, hopping is reduced and collisional effects are canceled
out.  It is difficult to see how to implement this kind of 
interference in a 2D or 3D configuration.

While collisional effects are considered here as an unwanted degrading
factor, our work also suggests the potential application of the
Ramsey separated field method for studying entanglement.

\begin{acknowledgments}
This work was supported in part by grants from the Minerva foundation
for a junior research group, the U.S.-Israel Binational Science
Foundation (grant No.~2002147), the Israel Science Foundation for a
Center of Excellence (grant No.~8006/03), and the German Federal
Ministry of Education and Research (BMBF) through the DIP project.
Useful conversations with Paul Julienne and Carl Williams are
gratefully acknowledged.
\end{acknowledgments}

\end{document}